\documentclass[11pt]{amsart}

\usepackage{amssymb}
\usepackage{amsmath,amsthm,amsfonts}  
 \usepackage{graphicx}
\usepackage{amssymb}
\usepackage{amsmath}
\usepackage{amsthm}
\usepackage{pdfsync}
\usepackage{fancyhdr}
\usepackage[applemac]{inputenc}
\usepackage{float}

\usepackage[default]{frcursive}
\usepackage[T1]{fontenc}

\usepackage{color}
\usepackage{xcolor}
\usepackage[breaklinks,colorlinks,backref]{hyperref}
\hypersetup{
    colorlinks, 
    linktoc=all, 
    linkcolor=red,  
  citecolor=hyptxt,
  urlcolor=blue}
  \hypersetup{
  citebordercolor=,
  filebordercolor=green,
  linkbordercolor=blue
}
\definecolor{hyptxt}{rgb}{0.7, 0.4, 0.9}
\usepackage{bibtopic}

\usepackage[full]{textcomp} 
     \usepackage{fbb} 
     \usepackage[scaled=.95]{cabin}
     \usepackage[varqu,varl]{inconsolata}
     \usepackage[libertine,bigdelims,vvarbb]{newtxmath}
     \usepackage[cal=boondoxo]{mathalfa}
     \usepackage[T1]{fontenc}
\useosf

\newtheorem{prop}{Proposition}[section]

\renewcommand{\qed}{\hfill $\square$}
\definecolor{hervecolor}{rgb}{0.8,0,0.7}
     
\newcommand{\ket}[1]{|\kern.3ex#1\kern.3ex\rangle}
\newcommand{\bra}[1]{\langle\kern.3ex #1 \kern.3ex|}
\newcommand{\scalar}[2]{\langle\kern.3ex #1 \kern.3ex|\kern.3ex#2\kern.3ex\rangle}

\newcommand{\ii}{\mathsf{i}}

\def\R{\mathbb{R}}

\def\C{\mathbb{C}}
\def\Z {\mathbb{Z}}

\def\deq{\stackrel{\mathrm{def}}{=}}

\numberwithin{equation}{section}

\begin{document}
\date{\today}
 
\title[$q$-triplets]{M\"{o}bius transforms, cycles  and $q$-triplets in statistical mechanics}
\author[J.-P. Gazeau and C. Tsallis]{
Jean-Pierre Gazeau$^{\mathrm{a,b}}$ and Constantino Tsallis$^{\mathrm{a,c,d}}$}

\address{\emph{$^{\mathrm{A}}$ Centro Brasileiro de Pesquisas F\'{\i}sicas } \\
\emph{Rua Xavier Sigaud 150, 22290-180 - Rio de Janeiro, RJ, Brazil  }} 
\address{\emph{  $^{\mathrm{B}}$ APC, UMR 7164}\\
\emph{Univ Paris  Diderot, Sorbonne Paris Cit\'e}  
\emph{75205 Paris, France}}
\address{\emph{  $^{\mathrm{C}}$ Santa Fe Institute}\\
\emph{1399 Hyde Park Road, New Mexico 87501, USA}} 
\address{\emph{  $^{\mathrm{D}}$ Complexity Science Hub Vienna}\\
\emph{Josefstaedter Strasse 39, A 1080 Vienna, Austria} 
} 

\email{e-mail:
gazeau@apc.univ-paris7.fr,tsallis@cbpf.br}

{\abstract{In the realm of Boltzmann-Gibbs (BG) statistical mechanics and its $q$-generalisation for complex systems, we analyse observed sequences of $q$-triplets, or $q$-doublets if one of them is the unity, in terms of cycles of successive M\"{o}bius transforms of the line preserving unity ( $q=1$ corresponds to the BG theory). Such transforms have the form $q\mapsto (aq + 1-a)/[(1+a)q -a]$, where $a$ is a real number; the particular cases $a=-1$ and $a=0$ yield respectively $q\mapsto (2-q)$ and $q\mapsto 1/q$, currently known as additive and multiplicative dualities. This approach seemingly enables the organisation of various complex phenomena into different classes, named $N$-complete or incomplete.  The classification that we propose here hopefully constitutes a useful guideline in the search, for non-BG systems whenever well described through $q$-indices, of new possibly observable physical properties.}}

\maketitle


\tableofcontents

\section{Introduction}
\label{intro}

Together with Maxwell electromagnetism, and classical, quantum and relativistic mechanics, Boltzmann-Gibbs (BG) statistical mechanics constitutes a pillar of contemporary theoretical physics. This powerful theory is based on the optimisation, under simple appropriate constraints, of the (additive) BG entropic functional $S_{BG}$, where $S_{BG}=k\sum_{i=1}^W p_i \ln (1/p_i)$ (with $\sum_{i=1}^Wp_i=1$, $k$ being a conventional positive constant chosen once and forever) for a simple discrete system with $W$ microscopic possibilities whose occurrence probabilities are given by $\{ p_i \}$. This fact typically leads, for a nonlinear (conservative or dissipative) dynamical classical system with {\it positive} maximal Lyapunov exponent,  to a (asymptotically) {\it linear} time growth of $S_{BG}(\{ p_i(t)\})$ while occupying a finely partitioned phase space. It also leads, for nearly all initial conditions for quite generic BG systems, to an {\it exponential} time relaxation towards its stationary state. Finally, this stationary state (usually referred to as {\it thermal equilibrium}) is characterised by the celebrated BG {\it exponential} weight $p_i=e^{-E_i/kT}/Z_{BG}$  with its partition function given by $Z_{BG} \equiv \sum_{j=1}^W e^{-E_j/kT}$, $E_j$ being the  energy eigenvalue  of the $j$-th state of a quantum conservative Hamiltonian system with specific boundary conditions.

For vast classes of complex natural, artificial and social systems, this relatively simple scenario fails. More precisely, either it is discrepant from experimental, observational or computational evidences, or it is plainly not calculable (i.e., mathematically ill-defined), typically because its partition function $Z_{BG}$ diverges, as already alerted long ago by Gibbs himself \cite{Gibbs1902}.  Consequently, a generalisation of the BG theory becomes mandatory. Such a generalisation has been proposed in 1988 \cite{Tsallis1988}, and has been useful for wide classes of complex systems, e.g., cold atoms \cite{coldatoms}, high-energy collisions of elementary particles \cite{particles}, granular matter \cite{granular}, low-dimensional maps \cite{standard}, asymptotically scale-free networks \cite{networks}, cosmic rays \cite{cosmic}, to quote but a few of them. This generalisation consists in optimising, under appropriate simple constraints, a nonadditive entropy $S_q$ through the introduction of a deformation parameter $q$, namely
\begin{equation}
S_q=k\sum_{i=1}^W p_i \ln_q \frac{1}{p_i}=-k\sum_{i=1}^W p_i \ln_{2-q} \frac{1}{p_i}\;\;\;(q \in \mathbb{R}\,; \,S_1=S_{BG}=-k\sum_{i=1}^W p_i \ln p_i) \,,
\end{equation}
with $\ln_q z \equiv \frac{z^{1-q}-1}{1-q} \;\;(\ln_1 z=\log z\, ; \, z \in \mathbb{C})$; its inverse function is given by $e_q^z=[1+ (1-q) z]^{1/(1-q)}$. The fact that $S_q$ is generically nonadditive is straightforwardly verified, more precisely as follows:
\begin{equation}
\frac{S_q(A+B)}{k}=\frac{S_q(A)}{k} + \frac{S_q(B)}{k} +(1-q) \frac{S_q(A)}{k}\frac{S_q(B)}{k}
\end{equation}
where $A$ and $B$ are {\it probabilistically independent}, i.e., $p_{ij}^{A+B}= p_i^A p_j^B, \, \forall (i,j)$. This property recovers, for $q=1$, the well known additivity of the $S_{BG}$ functional.

For wide classes of nonlinear dynamical systems with {\it zero}, instead of positive, maximal Lyapunov exponent, it turns out that the linear time growth occurs for $S_{q_{entropy}}$ with $q_{entropy}<1$. Concomitantly, relaxation occurs $q_{relaxation}$-exponentially with $q_{relaxation}>1$ towards a stationary state characterised by  $p_i=e_{q_{energy}}^{-E_i/kT}/Z_{q_{energy}}$  with the partition function given by $Z_{q_{energy}} \equiv \sum_{j=1}^W e_{q_{energy}}^{-E_j/kT}$ with $q_{energy} \ne 1$ \cite{Tsallis1988}. 

The situation is sometimes more complex than just described. For instance, the distribution of momenta of a many-body Hamiltonian system usually follows a $q_{moment}$-Gaussian form with $q_{moment} \ne 1$ not necessarily coincident with $q_{energy}$. The general scenario consists that, for a given complex system, we may have an infinite countable set of different $q$'s, corresponding to different one-body or many-body properties. However, only a small number of these $q$'s are in principle independent, all the others being related to those few through relatively simple analytic relations. The whole scenario appears to be strongly reminiscent of the scaling relations existing between the various exponents that emerge in the theory of critical phenomena (e.g., $\alpha+2\beta+\gamma=2$, $(2-\eta)\nu=\gamma$, $d\nu = 2-\alpha$, and similar ones). \\
The present work constitutes an attempt to formally establish, at least for some important classes of systems, the relations between the various $q$'s that are necessary to fully characterise the universality classes associated with a given complex system. In this attempt we follow along the lines of \cite{Tsallis2004,BurlagaVinas2005,TsallisGellMannSato2005,Tsallis2017a}. 
The needed algebraic and geometric material is described in Section \ref{geom} in terms of SU$(1,1)$ $\sim$ SL$(2,\R)$ homographic group actions on the unit disk (respectively upper half-plane) \cite{vilenkin68,gazeau09}.  In Section \ref{subset} we restrict these actions to particular ones leaving some point fixed, and study the corresponding subsets of the groups. In Section \ref{cycles} we proceed with the analysis of two-term or three-term cycles in view of displaying some universal relation when they involve a doublet or a triplet of parameters $q$. The same is carried out in Section \ref{var2} in terms of alternative variables. Numerical examples extracted from various observations are analysed  in Section \ref{num} from the point of view of such universality rules. Final discussion and comments constitute the content of  Section \ref{conclu}. 

\section{Unit disk, circle, half-plane, and real line: a reminder }
\label{geom}
Since specific conformal or homographic transformations of the real line occupy the central role in the present work, we think useful to give an overview   of its mathematical context. More details can be found in chapters VI and VII of \cite{vilenkin68} or in chapter 8 in \cite{gazeau09}.

The (open)  unit disk  in the complex plane is defined as 
\begin{equation}
\label{unitD}
\mathcal{D} \deq \{z  \in \C\, , \, \vert z \vert < 1\}\, .
\end{equation}  
Besides the unit disk, there is another equivalent representation commonly used in two-dimensional hyperbolic geometry, namely  the  Poincar\'e half-plane, and defining a model of hyperbolic space on the upper half-plane. The disk $\mathcal{D}$ and the upper half plane $\mathrm{P}_+ = \{ Z \in \C\, , \, \mathrm{Im} Z > 0\}$ are related by a conformal map, called   M\"{o}bius transformation,
\begin{equation}
\label{moebius}
\mathrm{P}_+ \ni Z \mapsto z = e^{\ii \phi}\, \frac{Z - Z_0}{Z - \overline{Z}_0} \in \mathcal{D}\, ,
\end{equation}
$\phi$ and $Z_0$ being arbitrary, and $\overline{Z}$ is the complex conjugate of $Z$. The canonical mapping is given by $Z_0 = \ii$ and $\phi = \pi/2$. It takes $\ii$ to the center of the disk and the origin $O$ to the bottom of the disk.

Like the sphere $S^2$ is invariant under space rotations forming the group \\
SO$(3) \simeq $SU$(2)/\Z_2$, $\Z_2 = \{ 1, -1\}$, the unit disk $\mathcal{D}$ is invariant under transformations of the \emph{homographic} or M\"{o}bius type:
\begin{equation}
\label{homsu11}
\mathcal{D} \ni z \mapsto z' = (\alpha \, z + \beta)\, (\bar\beta \, z + \bar \alpha)^{-1} \in \mathcal{D}\, , 
\end{equation}
with $\alpha$, $\beta \in \C$ and $\vert \alpha\vert^2 - \vert \beta\vert^2 \neq 0$. Since a common factor of $\alpha$ and $\beta$ is unimportant in the transformation (\ref{homsu11}), one can associate to the latter the $2\times 2$ complex matrix
\begin{equation}
\label{su11matrix}
\begin{pmatrix}
  \alpha    &  \beta  \\
  \bar\beta    &  \bar\alpha
\end{pmatrix} \deq g\, , \quad \mbox{with}\ \det{g} = \vert \alpha\vert^2 - \vert \beta\vert^2 = 1\, ,
\end{equation}
and we will write $z' = g\cdot z$. These matrices form the  group  SU$(1,1)$, one of the simplest examples of a simple, non-compact Lie group. It should be noticed that SU$(1,1)$ leaves invariant  the boundary $\mathbb{S}^1 \simeq $ U$(1)$ of $\mathcal{D}$ under the transformations (\ref{homsu11}).\\
Let us turn  our attention  to the corresponding symmetries in  the Poincar\'e half-plane. The conversion is carried out through a simple multiplication of matrices involving the specific (or ``canonical'') M\"{o}bius transformation,  written here as
\begin{equation}
\label{diskhalfplane3}
 z=\frac{Z-\ii}{1-\ii Z} \equiv \frac{1}{\sqrt{2}}\, \begin{pmatrix}
  1    &   -\ii \\
  -\ii    &  1
\end{pmatrix} \cdot Z \equiv {\tt m}\,\cdot Z\, , \quad Z = {\tt m}^{-1}\cdot z\, , 
\end{equation}
and conversely
\begin{equation}
\label{halplanedisk}
Z=\frac{z+\ii}{\ii z + 1}\, . 
\end{equation}
Note that when extended to the boundaries, the bijection \ref{diskhalfplane3} is a  Cayley transformation that maps in a stereographic way  the unit circle  $\mathbb{S}^1$ onto the real line
\begin{equation} 
\label{StoR}
\mathbb{S}^1 \ni  e^{\ii\theta} \mapsto t=\frac{e^{\ii\theta}+\ii}{\ii \,e^{\ii\theta}+1}\in \R\, , \  \theta \in [0, 2\pi)\,  ,
 \end{equation}
where $\theta=0  \mapsto t=1$, $\theta=\pi/2 \mapsto t=\infty$, $\theta=\pi \mapsto t=-1$ and $\theta=\frac{3\pi}{2} \mapsto 0$.
Conversely, 
\begin{equation}
\label{RtoS}
\R \ni  t\mapsto e^{\ii\theta} =\frac{t-\ii}{- \ii \,t +1}\in \mathbb{S}^1\,.
\end{equation}
Therefore, the transformation $z' = g\cdot z$ where $g \in $ SU$(1,1)$ becomes in the half-plane 
\begin{equation}
\label{ZtoZp}
Z' = s\cdot Z = \frac{a Z+b}{cZ+d}\, , 
\end{equation}
with 
\begin{equation}
\label{passgs}
s = {\tt m}^{-1}\, g \, {\tt m}= \begin{pmatrix}
\mathrm{Re} \alpha + \mathrm{Im} \beta      &  \mathrm{Im}  \alpha +  \mathrm{Re} \beta  \\
- \mathrm{Im}  \alpha +  \mathrm{Re} \beta   &  \mathrm{Re} \alpha - \mathrm{Im} \beta 
\end{pmatrix} \equiv \begin{pmatrix}
   a   & b   \\
  c   &  d
\end{pmatrix}\, , \quad a, \, b\, , c\, , d\, \in \R\, ,
\end{equation}
and conversely, 
\begin{equation}
\label{ipassgs}
g = {\tt m}\, s \, {\tt m}^{-1}=\frac{1}{2} \begin{pmatrix}
  a+d +\ii (b-c)    & b+c +\ii(a-d)  \\
 b+c -\ii (a-d)  &   a+d -\ii (b-c) 
\end{pmatrix} \equiv \begin{pmatrix}
   \alpha   & \beta   \\
  \bar\beta   &  \bar\alpha
\end{pmatrix}\, , \quad \alpha, \, \beta \in \C\, .
\end{equation}
Since $\det s = 1$, the set of such $2\times 2$ real matrices form the group  SL$(2,\R)$, which leaves invariant the upper half-plane and its boundary, which is the real line $\R$.   

Indeed, the action (\ref{homsu11}) of $g = \begin{pmatrix}
 \alpha     &   \beta \\
   \bar \beta   &  \bar \alpha
\end{pmatrix}\in  \mbox{SU}(1,1)$ on $\mathcal{D}$ extends to the boundary as
\begin{equation}
\label{su11actbound8}
g\cdot e^{\ii \theta} = (\alpha\,  e^{\ii \theta} + \beta)\, (\bar\beta\, e^{\ii \theta}  + \bar \alpha)^{-1} \equiv e^{\ii \theta'}\in \mathbb{S}^1\, , 
\end{equation}
  and so leaves the latter invariant. Similarly, SL$(2,\R)$ acts homographically on the real line $\R$ as 
  \begin{equation}
\label{xtoxp}
\R \ni q \mapsto s\cdot q= \frac{aq+b}{cq+d} = q^{\prime}\in \R\, .  
\end{equation}
Below, we extend SL$(2,\R)$ homographic transformations of the line, \eqref{xtoxp}, to   SL$^{\pm}(2,\R)$,  the not connected group of real $2 \times 2$ matrices with determinant $\pm 1$, in order to include the simple  inversion $i_v$
\begin{equation}
\label{inv}
q\mapsto 1/q= \begin{pmatrix}
   0   &  1  \\
   1   &  0
\end{pmatrix}\cdot x := i_v\cdot x \, .
\end{equation}
Similarly,  we extend SU$(1,1)$ homographic transformations of the circle, \eqref{su11actbound8}, to SU$^{\pm}(1,1)$ the group of  matrices with determinant $\pm 1$. We note that the image of \eqref{inv} under the Cayley transform \eqref{ipassgs} is the same matrix 
\begin{equation}
\label{iv}
i_v=  \begin{pmatrix}
   0   &  1  \\
   1   &  0
\end{pmatrix}\,. 
\end{equation}

\section{A subset of M\"{o}bius transformations}  
 \label{subset}
 We now examine the subset $\mathcal{A}$ of SL$^{\pm}(2,\R)$ made of elements obeying the two constraints:
 \begin{enumerate}
  \item[(i)] They are nilpotent,
  \begin{equation}
\label{nilpot}
s^2 = I\, . 
\end{equation}
  \item[(ii)] They leave $x=1$ invariant under  M\"{o}bius transformations,
\begin{equation}
\label{Iinv}
s\cdot 1 = 1\, . 
\end{equation}
\end{enumerate}
 \begin{prop}
 The subset  $\mathcal{A}$ of SL$^{\pm}(2,\R)$ made of elements obeying (i) and (ii) contains the identity $I$ (up to a sign) and  the following one parameter  family of $2\times 2$ real matrices 
 \begin{equation}
\label{salpha}
\ell(a) = \begin{pmatrix}
  a  & 1-  a \\
 1+ a     &  -a
\end{pmatrix}\, , \quad a \in \R\, . 
\end{equation}
\end{prop}
\proof
Let us start with a generic element $s= \begin{pmatrix}
  a    &  b  \\
   c   &  d
\end{pmatrix}$ in the algebra M$_2(\R)$ with arbitrary determinant $\Delta = ad-bc$. Nilpotence  and condition that $1$ is a fixed point under $s\cdot q= (aq+b)(cq+d)^{-1}$ entail the following equations on the matrix elements
\begin{align}
\label{cond1}
  a^2 + bc &= 1 \, ,    \\
 \label{cond2}  d^2 + bc  &= 1  \, , \\
\label{cond3}   (a+d)b&=0 = (a+d)c\, \\
   \label{cond4} a+b &= c+d \,.
\end{align}
The condition $ \Delta = \pm 1$ results from the above, of course. 
\eqref{cond1} and \eqref{cond2} imply  $a^2=d^2$, and so we have the equivalent conditions
\begin{align}
\label{cond1p}
  a^2 + bc &= 1 \, ,    \\
 \label{cond2p} a  &= \pm d  \, , \\
\label{cond3p}   (a+d)b&=0 = (a+d)c\, \\
   \label{cond4p} a+b &= c+d \,.
\end{align}
\subsubsection*{Case $a=d=0$}
Then, $b=c$ (from \eqref{cond4p}), and $bc=b^2=1$ (from \eqref{cond1p}). Thus, $s=\pm i_v$.
\subsubsection*{Case $a=d \neq 0$}
Then, $b=0=c$ (from \eqref{cond3p}), and $a=\pm 1= d$ from \eqref{cond1p}). Thus, $s=\pm I$.
\subsubsection*{Case $a=-d \neq 0$}
Then $bc= 1-a^2$ (from \eqref{cond1p}) and $c-b = 2a$ (from \eqref{cond4p}), which means that $c$ and $-b$ are roots of $X^2 -2a X +a^2 -1= 0$, i.e, $b=1-a$ and $c=1+a$, or $b=-1-a$ and $c=a-1$. 
These two possibilities yield
\begin{equation*}
s= \begin{pmatrix}
  a    &  1-a  \\
  1+a    & -a 
\end{pmatrix}= \ell(a) \, ,\quad \mbox{or} \quad s= \begin{pmatrix}
  a    &  -1-a  \\
  -1+a    & -a 
\end{pmatrix} =- \ell(-a) \, .
\end{equation*}
Since the second one is equivalent, as a M\"{o}bius transformation of $\R$, to $\ell(-a)$ , and that $a\in \R$, it results that all possible solutions are 
$\ell(a)$ with $a \in \R$, together with the identity $I$, and up to a factor  $\neq 0$. 
\qed

Consequently, from now on we focus on the particular cases of the map \eqref{xtoxp}, namely
\begin{equation}
\label{xtoxpla}
\R \ni q \mapsto l(a)\cdot q= \frac{aq+1-a}{(1+a)q-a}\, .  
\end{equation}

Elements of $\mathcal{A}$ have interesting properties. 
\begin{enumerate}
  \item[$\mathcal{P}_1$:] Determinant 
  \begin{equation}
\label{detA}
\det \ell(a) = -1 \quad \forall a \in \R\,. 
\end{equation}
Thus all elements of $\mathcal{A}$, except the identity $I$, have determinant equal to $-1$. 
 \item[$\mathcal{P}_2$:] Inverse (from nilpotence)
  \begin{equation}
\label{invla}
\ell^{-1}(a) = \ell(a)  \quad \forall a \in \R\,.
\end{equation}
\item[$\mathcal{P}_3$:] Parameter inversion 
  \begin{equation}
\label{invla2}
\ell(-a)= i_v\, \ell(a)\,i_v \,, \ \Leftrightarrow \ i_v \, \ell(-a)=  \ell(a)\,i_v\, .
\end{equation}
 \item[$\mathcal{P}_4$:] 
 Particular cases: inversion, affine transformation of the line, and their combination,
 \begin{equation}
\label{parcas}
\ell(0) = i_v\, , \quad  \ell(-1)= \begin{pmatrix}
    -1  &  2  \\
     0 &   1
\end{pmatrix} \, , \quad \ell(1)= \begin{pmatrix}
    1  &  0  \\
     2 &   -1
\end{pmatrix}= i_v \, \ell(-1) \, i_v \, .
\end{equation}
 \item[$\mathcal{P}_5$:] Composition is internal up to inversion
  \begin{equation}
\label{lalap}
 \ell(a) \,  \ell(a^{\prime})= \begin{pmatrix}
  1+ a^{\prime}-a    &  a-a^{\prime} \\
  a^{\prime}-a    & 1 +  a-a^{\prime}
\end{pmatrix}= i_v  \ell(a^{\prime}-a)\, . 
 \end{equation}
 \end{enumerate}
Hence, since the product of two arbitrary  distinct elements and different of the identity has determinant equal to $1$,  $\mathcal{A}$  is not a subgroup of SL$^{\pm}(2,\R)$. On the other hand, the subset 
\begin{equation}
\label{ivA}
i_v\mathcal{A}= \left\lbrace i_v\ell(a)\, , \, a \in \R\right\rbrace
\end{equation}
is an abelian subgroup isomorphic to $\R$. Indeed, it contains $I$, and \eqref{invla} and \eqref{lalap} imply
\begin{equation}
\label{ivlivl}
i_v\ell(a)i_v\ell(a^{\prime}) = i_v\ell(a + a^{\prime})\, \Leftrightarrow \ell(a)\ell(a^{\prime})= \ell(a -a^{\prime})i_v \, , 
\end{equation}
 which can be directly verified with 
 \begin{equation}
\label{ivl}
i_v\ell(a)= \begin{pmatrix}
  1+a    &  -a  \\
    a  &  1-a
\end{pmatrix} = I + a\varpi\, , \quad \varpi = \begin{pmatrix}
   1   &  -1  \\
    1  &  -1
\end{pmatrix}\, , \quad \varpi^2= 0\, . 
\end{equation}

Let us now write down the counterpart  $\mathcal{B}$ of $\mathcal{A}$ as a subset of SU$^{\pm}(1,1)$ by using  \eqref{ipassgs}:
\begin{equation}
\label{defB}
\mathcal{B}= \left\lbrace  \jmath (a) = \begin{pmatrix}
 -\ii a     & 1 + \ii a  \\
  1 - \ii a    &  \ii a
\end{pmatrix}\, , \, a \in \R \right\rbrace\, . 
\end{equation}
As a nilpotent homographic transformation of the unit circle, it leaves invariant the point $1$. 

\section{Cycles}
\label{cycles}
We now examine specific sequences of maps \eqref{xtoxpla} in view of their   relevance to  relations between   parameters $q$ associated with  different facets of a complex system.
\subsection{Two-term cycles}

Let us consider the two-term cycle 
\begin{equation}
\label{2termcycle}
q_1\mapsto q_2=\ell(a_{12})\cdot q_1\mapsto q_1=\ell(a_{21})\cdot q_2\, ,  
\end{equation}
which reads after using \eqref{lalap} and \eqref{invla2},
\begin{equation}
\label{2x1x1}
q_1= i_v\ell(a_{12} - a_{21})\cdot q_1\, . 
\end{equation}
This leads to consider the algebraic relation between $a =   a_{12} - a_{21} $ and its fixed point $q=i_v\ell(a)\cdot q$, i.e., $(1+a)q^2 -(1+2a) q + a=0$. This equation 
has two solutions, $q=1$ (expected), and if $a \neq -1$, $q= a /(a +1)$, i.e., $a = q/(1-q)$. Thus, for arbitrary $q_1\neq 1$, \eqref{2x1x1} does provide non trivial solutions for the doublet $a_{12}$,   and  $a_{21}$, depending on the initial $q_1$, 
\begin{equation}
\label{ntsol2}
a_{12} - a_{21} = \frac{q_1}{1-q_1}=  \frac{1}{1-q_1} -1\, . 
\end{equation}
Let us introduce for our next purposes the quantities 
\begin{equation}
\label{defaux2}
q_{\mathrm{aux}}:= q_1 \quad\mbox{and}\quad  a_{\mathrm{aux}}:=\frac{1}{q_{\mathrm{aux}}-1}
\end{equation} 
which are determined by the fixed point. We then get from the above the ``conservation law":
\begin{equation}
\label{conslaw2}
a_{12} - a_{21} + a_{\mathrm{aux}}= -1\,. 
\end{equation}

\subsection{Three-term cycles}

Let us now consider the three-term cycle 
\begin{equation}
\label{3termcycle}
q_1\mapsto q_2=\ell(a_{12})\cdot q_1\mapsto q_3=\ell(a_{23})\cdot q_2\mapsto q_1=\ell(a_{31})\cdot q_3\, ,  
\end{equation}
which reads after using \eqref{lalap} and \eqref{invla},
\begin{equation}
\label{x1x1}
q_1= \ell(a_{31} - a_{23} + a_{12})\cdot q_1\, . 
\end{equation}
This leads to consider the algebraic relation between $a =   a_{31} - a_{23} + a_{12}$ and its fixed point $q=\ell(a)\cdot q$, i.e., $(1+a)q^2 -2a q + a-1=0$. This equation 
has two solutions, $q=1$ (expected), and if $a \neq -1$, $q= (a -1)/(a +1)$, i.e., $a = (1+q)/(1-q)$. Thus, for arbitrary $q_1\neq 1$, \eqref{x1x1} does provide non trivial solutions for the triplet $a_{31}$,  $a_{23}$, and  $a_{12}$, depending on the initial $q_1$, 
\begin{equation}
\label{ntsol}
a_{31} - a_{23} + a_{12}= \frac{1+q_1}{1-q_1}=  \frac{1}{1-(1+q_1)/2} -1\, . 
\end{equation}
Like above,  we introduce  the quantities
\begin{equation}
\label{defaux3}
q_{\mathrm{aux}}:= \frac{1+q_1}{2} \quad\mbox{and}\quad  a_{\mathrm{aux}}:=\frac{1}{1-q_{\mathrm{aux}}}
\end{equation} 
which are determined by the fixed point. Note the opposite sign of the latter with regard to the previous case. There results the three-term conservation law:
\begin{equation}
\label{conslaw}
a_{31} - a_{23} + a_{12} - a_{\mathrm{aux}}= -1\,. 
\end{equation}

\subsection{$N$-term cycles}
The two above cases allow us to easily infer the general $N$-term case. If $N=2p$ is even, Eq. \eqref{2x1x1} generalizes to
 \begin{equation}
\label{2px1x1}
q_1= i_v\ell\left(\sum_{i=1}^{N-1}(-1)^{i+1}a_{i i+1} - a_{N1}\right)\cdot q_1\, , 
\end{equation}
and yields the fixed point  $q= a /(a +1)$, i.e., $a = q/(1-q)$, with $a = \sum_{i=1}^{N-1}(-1)^{i+1}a_{i i+1} - a_{N1}$, and the resulting conservation law 
\begin{equation}
\label{conslaw2p}
\sum_{i=1}^{N-1}(-1)^{i+1}a_{i i+1} - a_{N1} + a_{\mathrm{aux}}= -1\, , 
\end{equation}
where $a_{\mathrm{aux}}:= \dfrac{1}{q_{\mathrm{aux}}-1}$, $q_{\mathrm{aux}}:= q_1$. 
 
 If $N=2p+1$ is odd, Eq. \eqref{x1x1} generalizes to
 \begin{equation}
\label{2p+1x1x1}
q_1= i_v\ell\left(\sum_{i=1}^{N-1}(-1)^{i+1}a_{i i+1} + a_{N1}\right)\cdot q_1\, , 
\end{equation}
and yields the fixed point  $q= (a-1) /(a +1)$, i.e., $a = (1+q)/(1-q)$, with $a = \sum_{i=1}^{N-1}(-1)^{i+1}a_{i i+1} + a_{N1}$, and the resulting conservation law 
\begin{equation}
\label{conslaw2p2}
\sum_{i=1}^{N-1}(-1)^{i+1}a_{i i+1} + a_{N1} - a_{\mathrm{aux}}= -1\, , 
\end{equation}
where $a_{\mathrm{aux}}:= \dfrac{1}{1-q_{\mathrm{aux}}}$, $q_{\mathrm{aux}}:=(1+q_1)/2$. 
 
Let us finally remark that the extension of these formulas and relations to the case of complex variables is straightforward. The underlying group is SL$^{\pm}(2,\C)$, the group of conformal transformations of the complex plane or of the (Riemann) sphere. 

\section{With another variable}
\label{var2}
In the previous section, besides the nilpotence, we imposed that a finite point, namely $1$, be left invariant. It is instructive to impose now that the point at the infinity be left invariant. A conformal transformation which sends $1$ to $\infty$ is given by
\begin{equation}
\label{xtoy}
q \mapsto \tilde{q}= \frac{\lambda}{1-q} = \begin{pmatrix}
 0     &  \lambda  \\
 - 1    &   1
\end{pmatrix}\cdot q \Leftrightarrow q = \frac{1}{\lambda}\begin{pmatrix}
  1   & - \lambda  \\
   1   &  0
\end{pmatrix}\cdot \tilde{q}\,. 
\end{equation}
where  $\lambda >0$ is a parameter. The M\"{o}bius transformation $\ell(a)$, \eqref{salpha} , becomes:
\begin{align}
\label{ellto} \nonumber
\tilde{q}^{\prime}&= \begin{pmatrix}
 0     &  \lambda  \\
  -1    &   1
\end{pmatrix}\cdot q^{\prime} = \frac{1}{\lambda} \begin{pmatrix}
 0     &  \lambda  \\
 - 1    &   1
\end{pmatrix} \begin{pmatrix}
  a   & 1-  a \\
 1+ a     &  -a
\end{pmatrix} 
\begin{pmatrix}
  1    &  -\lambda  \\
   1   &  0
\end{pmatrix}\cdot \tilde{q} \\
&= 
\begin{pmatrix}
   1   &  -\lambda(a +1)   \\
   0   &  -1
\end{pmatrix}\cdot \tilde{q} = \begin{pmatrix}
  - 1   &  \lambda(a +1)   \\
   0   &  1
\end{pmatrix}\cdot \tilde{q} \equiv m_{\lambda}(a+1)\cdot \widetilde{q}\, .
\end{align}
Thus, the corresponding M\"{o}bius transformation reduces to a translation combined with the space inversion $\tilde{q}\mapsto -\tilde{q} = i_s\cdot \tilde{q}$. By introducing the abelian group $\mathcal{T}_{\lambda}$ of translations of the real line, 
\begin{equation}
\label{Tlb}
\mathcal{T}_{\lambda}= \left\lbrace \mathsf{t}_{\lambda} (b) = \begin{pmatrix}
   1   &  \lambda b  \\
    0  &  1
\end{pmatrix} \right\rbrace\, , \quad \mathsf{t}_{\lambda} (b) \mathsf{t}_{\lambda} (b^{\prime}) =\mathsf{t}_{\lambda} (b + b^{\prime}) \, , 
\end{equation}
the new transformations read
\begin{equation}
\label{newtr}
m_{\lambda}(b) = \mathsf{t}_{\lambda} (b) \,i_s = i_s\,  \mathsf{t}_{\lambda} (-b)\, , \quad  m_{\lambda}(b) \,m_{\lambda}(b^{\prime}) =  m_{\lambda}(b -b^{\prime})\, i_s\, .  
\end{equation}
and have  similar properties to the $\ell(a)$'s above.  

Despite the simpler nature of the above geometric operations,  we choose in this paper to work with the previous formalism established from the fixed point $q=1$.  Indeed, this value corresponds, in the present context, to the BG particular instance.

\section{Observational examples}
\label{num}
In this section, we list a series of  three-term and two-term cycles issued from various observations.  

The first experimental evidence of the existence of a $q$-triplet in nature, conjectured in \cite{Tsallis2004}, was achieved in the magnetic fluctuations of the solar plasma as measured at the Voyager 1 near the end of our planetary system \cite{BurlagaVinas2005}. Our present study is  based on  this  observation.
Three deformation parameters, $q_{\mathrm{sensitivity}}\equiv q_{\mathrm{sens}}$, $q_{\mathrm{stationarity}}\equiv q_{\mathrm{stat}}$, $q_{\mathrm{relaxation}}\equiv q_{\mathrm{rel}}$ are supposed to be part  of a  three-term cycle of the type described in \eqref{3termcycle} through the relations
\begin{equation}
\label{defal}
 \frac{1}{q_{\mathrm{sens}}-1} =:-a_{\mathrm{sens}}\, \quad \frac{1}{q_{\mathrm{stat}}-1} 
=: a_{\mathrm{stat}}\, \quad  \frac{1}{q_{\mathrm{rel}}-1} =:   a_{\mathrm{rel}}\,.
\end{equation}
This is precisely the guideline in the identification of the parameters $a_{23}, a_{31}, a_{12}$ in \eqref{3termcycle} with these $a_{sen}, a_{stat}, a_{rel}$ respectively.
Our objective is to reveal  a kind of regularity in the sequence of fixed points, or equivalently, of $q_{\mathrm{aux}}$'s, as defined in \eqref{defaux3}.
Furthermore, the two-term cycles are considered if one of the three deformation parameters is equal to $1$, which corresponds to the BG statistics.  All numerical data are summarised in Table \ref{qalpha}.

\subsection{Observations with three-term cycles}
\subsubsection{Solar wind}

The following conjectural values are from \cite{TsallisGellMannSato2005}
\begin{eqnarray}
q_{\mathrm{sens}} &=& -1/2  \label{setq1}\\
q_{\mathrm{stat}} &=&  7/4   \label{setq2} \\
q_{\mathrm{rel}} &=& 4
\label{setq3}
\end{eqnarray}
hence
\begin{eqnarray}
a_{\mathrm{sens}}:= \frac{1}{1-q_{\mathrm{sens}}} &=& 2/3  \\
a_{\mathrm{stat}}:= \frac{1}{q_{\mathrm{stat}}-1} &=&  4/3    \\
a_{\mathrm{rel}}:=\frac{1}{q_{\mathrm{rel}}-1} &=& 1/3 
\end{eqnarray} 
hence
\begin{eqnarray}
\frac{1}{q_{\mathrm{rel}}-1} &=&  \frac{1}{q_{\mathrm{sens}}-1}+1   \\
\frac{1}{q_{\mathrm{stat}}-1} &=&   \frac{1}{q_{\mathrm{sens}}-1}+2 
\end{eqnarray} \\

One also checks, in order to follow the three-cycle relation \eqref{ntsol}, 
\begin{equation}
\label{sowEq}
a_{\mathrm{rel}} + a_{\mathrm{stat}} - a_{\mathrm{sens}} = 1\, , 
\end{equation}
which implies that the fixed point in this case is $q_1 = 0$, and $q_{\mathrm{aux}}= 1/2$.  

Let us incidentally mention a remarkable relation \cite{Baella2008}. If we define $\epsilon \equiv 1-q$, Eqs. (\ref{setq1}-\ref{setq3}) are equivalent to
\begin{eqnarray}
\epsilon_{\mathrm{sens}} &=& 3/2  \label{setepsilon1}\\
\epsilon_{\mathrm{stat}} &=&  -3/4   \label{setepsilon2} \\
\epsilon_{\mathrm{rel}} &=& -3 \,.
\label{setepsilon3}
\end{eqnarray}
We then verify
\begin{eqnarray}
\epsilon_{\mathrm{stat}} &=& \frac{\epsilon_{\mathrm{sens}}+\epsilon_{\mathrm{rel}}}{2} \;\;\;\;\;\mbox{(arithmetic mean)} \label{setmean1}\\
\epsilon_{\mathrm{sens}} &=& [ \epsilon_{\mathrm{stat}} \; \epsilon_{\mathrm{rel}}]^{1/2} \;\;\;\mbox{(geometric  mean)} \label{setmean2} \\
\epsilon_{\mathrm{rel}}^{-1} &=& \frac{\epsilon_{\mathrm{stat}}^{-1} + \epsilon_{\mathrm{sens}}^{-1}}{2} \;\;\;\;\mbox{(harmonic mean)} \,.
\label{setmean3}
\end{eqnarray}
The possible interpretation of these intriguing relations in terms of some special symmetry, or some analogous property, has proved elusive.

\subsubsection{Feigenbaum point}
From \cite{TsallisPlastinoZheng1997,RobledoMoyano2008,TsallisTirnakli2010}.

$q_{\mathrm{sens}}$ in \cite{TsallisPlastinoZheng1997}



$q_{\mathrm{rel}}$ in \cite{RobledoMoyano2008}

$q_{\mathrm{stat}}$ in \cite{TsallisTirnakli2010}

\begin{eqnarray}
q_{\mathrm{sens}} &=& 0.2445 \\
q_{\mathrm{stat}} &=&  1.65    \\
q_{\mathrm{rel}} &=& 2.2498
\end{eqnarray}
hence
\begin{eqnarray}
\frac{1}{1-q_{\mathrm{sens}}} &=& 1.3236 \\
\frac{1}{q_{\mathrm{stat}}-1} &=&  1.5385    \\
\frac{1}{q_{\mathrm{rel}}-1} &=& 0.8001
\end{eqnarray} 

Also, within the error bars, we verify that \cite{Baella2010}
\begin{equation}
\epsilon_{rel}+\epsilon_{sen}  \simeq \epsilon_{sen}\epsilon_{stat} \,,
\end{equation}
hence
\begin{equation}
q_{\mathrm{rel}}+q_{\mathrm{sens}} -2 \simeq [1-q_{\mathrm{sens}}] [q_{\mathrm{stat}}-1] \,.
\end{equation} 
Indeed, $[q_{\mathrm{rel}}+q_{\mathrm{sens}} -2] / [1-q_{\mathrm{sens}}] [q_{\mathrm{stat}}-1]=1.0066$.
Notice, however, that this relation does not belong to the set of those that we are discussing in the present paper.

We check in this case that 
\begin{equation}
\label{sowEq2}
a_{\mathrm{rel}} + a_{\mathrm{stat}} - a_{\mathrm{sens}} = 0.8001 + 1.5385 - 1.3236 = 1.015\, , 
\end{equation}
which implies that the fixed point in this case is $q_1 \approx 0.0075$ and $q_{\mathrm{aux}} \approx 1/2$.
 
\subsubsection{Brazos river}
From \cite{StosicStosicSingh2018}.

\begin{eqnarray}
q_{\mathrm{sens}} &=& 0.244 \\
q_{\mathrm{stat}} &=&  1.65    \\
q_{\mathrm{rel}} &=& 2.25
\end{eqnarray}
hence
\begin{eqnarray}
a_{\mathrm{sens}}  = \frac{1}{1-q_{\mathrm{sens}}} &=& 1.323 \\
 a_{\mathrm{stat}} =\frac{1}{q_{\mathrm{stat}}-1} &=&  1.538    \\
a_{\mathrm{rel}} = \frac{1}{q_{\mathrm{rel}}-1} &=& 0.8696
\end{eqnarray} 
In this case we have
\begin{equation}
\label{sowEq5}
a_{\mathrm{rel}} + a_{\mathrm{stat}} - a_{\mathrm{sens}} = 0.8696 + 1.538 - 1.323 = 1.0846\, , 
\end{equation}
which implies that the fixed point in this case is $q_1 \approx 0.0406$ and $q_{\mathrm{aux}} \approx 0.5203$.

\subsubsection{Bitcoin}
From \cite{StosicStosicLudermirStosic2018}.

\begin{eqnarray}
q_{\mathrm{sens}} &=& 0.14 \\
q_{\mathrm{stat}} &=&  1.54    \\
q_{\mathrm{rel}} &=& 2.25
\end{eqnarray}
hence
\begin{eqnarray}
a_{\mathrm{sens}}  = \frac{1}{1-q_{\mathrm{sens}}} &=& 1.163 \\
 a_{\mathrm{stat}} =\frac{1}{q_{\mathrm{stat}}-1} &=&  1.85    \\
a_{\mathrm{rel}} = \frac{1}{q_{\mathrm{rel}}-1} &=& 0.8696
\end{eqnarray} 
In this case we have
\begin{equation}
\label{sowEq4}
a_{\mathrm{rel}} + a_{\mathrm{stat}} - a_{\mathrm{sens}} = 0.8696 + 1.85 - 1.163 = 1.5566\, , 
\end{equation}
which implies that the fixed point in this case is $q_1 \approx 0.2176$ and $q_{\mathrm{aux}} \approx 0.6088$.

\subsubsection{Standard map}
From \cite{TirnakliBorges2016,RuizTirnakliBorgesTsallis2017}.
\begin{eqnarray}
q_{\mathrm{sens}} &=& 0 \\
q_{\mathrm{stat}} &=&  1.935    \\
q_{\mathrm{rel}} &=& 1.4
\end{eqnarray}
hence
\begin{eqnarray}
a_{\mathrm{sens}}  = \frac{1}{1-q_{\mathrm{sens}}} &=& 1 \\
 a_{\mathrm{stat}} =\frac{1}{q_{\mathrm{stat}}-1} &=&  1.0695    \\
a_{\mathrm{rel}} = \frac{1}{q_{\mathrm{rel}}-1} &=& 2.5
\end{eqnarray} 
In this case we have
\begin{equation}
\label{sowEq6}
a_{\mathrm{rel}} + a_{\mathrm{stat}} - a_{\mathrm{sens}} = 2.5 + 1.0695 - 1 = 2.5695\, , 
\end{equation}
which implies that the fixed point in this case is $q_1 \approx 0.4397$ and $q_{\mathrm{aux}} \approx 0.71985$

\subsubsection{Ozone layer}
From \cite{FerriSavioPlastino2010}.

\begin{eqnarray}
q_{\mathrm{sens}} &=& -8.1 \\
q_{\mathrm{stat}} &=&  1.32    \\
q_{\mathrm{rel}} &=& 1.89
\end{eqnarray}
hence
\begin{eqnarray}
a_{\mathrm{sens}}  = \frac{1}{1-q_{\mathrm{sens}}} &=& 0.11 \\
 a_{\mathrm{stat}} =\frac{1}{q_{\mathrm{stat}}-1} &=&  3.125    \\
a_{\mathrm{rel}} = \frac{1}{q_{\mathrm{rel}}-1} &=& 1.12
\end{eqnarray} 

In this case we have
\begin{equation}
\label{sowEq3}
a_{\mathrm{rel}} + a_{\mathrm{stat}} - a_{\mathrm{sens}} = 1.12 + 3.125 - 0.11 = 4.135\, , 
\end{equation}
which implies that the fixed point in this case is $q_1 \approx 0.61$ and $q_{\mathrm{aux}} \approx 0.8$.

\subsection{Observations with two-term cycles}

These cycles are degenerate three-term cycles in which one of the $q$'s is $=1$. 
For example, for 
$\tau = 1$ day we found that the three solar activity indices daily Sunspot Number (SN) from the Sunspot Index Data Center, Magnetic Field (MF) strength from the National Solar Observatory / Kitt Peak, and Total Solar Irradiance (TSI) means from Virgo/SoHO, may be essentially described by the $q$-triplet sets \cite{FreitasMedeiros2009}
$(q_{\mathrm{stat}}, q_{\mathrm{sens}},q_{\mathrm{rel}}) 
= (1.31 \pm 0.07, -0.71 \pm 0.10, 1), (1.21 \pm 0.06, -0.44 \pm 0.07, 1)$ and $(1.54 \pm 0.03, -0.52 \pm 0.10, 1)
$ respectively.

\begin{table}[H]
  \centering 
  \begin{tabular}{|l|c|c|c|c|c|}
\hline
  & $q_{\mathrm{sens}}$&$q_{\mathrm{stat}}$& $q_{\mathrm{rel}}$& $q_{\mathrm{aux}}$&$q_1$  \\
  \hline
Solar wind (conjectural)  & -1/2  & 7/4 & 4&0.5 &0\\
\hline
Solar wind (observations)  & $-0.6 \pm 0.2$  & $1.75 \pm 0.06$ & $3.8 \pm 0.3$&0.5158 &0.0316\\
\hline 
Feigenbaum point (calculations) & 0.2444877... &$1.65 \pm 0.05$  &2.2497841 & 0.50375& 0.0075\\
 \hline
Brazos river (observations) & 0.244 & 1.65   &2.25 &0.5203 &0.0406\\     
 \hline
Bitcoin (observations) & 0.14 & 1.54&2.25 &0.6088 & 0.2176\\
 \hline
 Standard map (calculations) & 0 & 1.935   &1.4 &0.71985 &0.4397\\
 \hline 
Ozone layer (observations)& -8.1 & 1.32 &1.89 &0.805  & 0.61\\ 
\hline
Solar activity/SN (observations)&$-0.71 \pm 0.10$ & $1.31 \pm 0.07$ & 1& 0.725&0.725\\
 \hline
Solar activity/MF (observations) & $-0.44 \pm 0.07$&$1.21 \pm 0.06$ & 1 & 0.803&0.803\\
 \hline
Solar activity/TSI (observations)& $-0.52 \pm 0.10$& $1.54 \pm 0.03$& 1&0.544 &0.544\\
\hline
\end{tabular}
  \caption{Numerical data from a non exhaustive series of observations displaying sensitivity, stationarity,  and relaxation $q$-indices, and their respective auxiliary indices issued from three-term cycle or two-term cycle fixed points $q_1$. For the three-term cycles, One can observe an interesting closeness between ``Solar wind'' \cite{BurlagaVinas2005,TsallisGellMannSato2005}, ``Feigenbaum point'' \cite{TsallisPlastinoZheng1997,TsallisTirnakli2010}, and ``Brazos river'' \cite{StosicStosicSingh2018} in the sense that, for all of them, $q_1\simeq 0$, whilst ``Bitcoin"\cite{StosicStosicLudermirStosic2018}, ``Standard map" \cite{TirnakliBorges2016,RuizTirnakliBorgesTsallis2017}, and ``Ozone layer"  \cite{FerriSavioPlastino2010},  are neatly apart from them. With those, as well as with the present two-term cycles, one cannot conclude.}\label{qalpha}
\end{table}

\section{Conclusion}
\label{conclu}

We have examined a (non exhaustive) series of observational data giving three types of deformation parameter $q$, namely $q_{\mathrm{sensitivity}}$, $q_{\mathrm{stationarity}}$, and $q_{\mathrm{relaxation}}$ by supposing they are  part  of an affine   three-term cycle, or of a two-term cycle if one of them is $1$. Our aim was to establish a kind of conservation law allowing to group the observed phenomenons into equivalence classes. In view of our results, one could conjecture that in the case of effective three-term cycles there exists a class for which the fixed point is $q_1 \approx 0$. 
For those cases, we might conjecture that this class of systems with $q_1=0$ has only two independent $q$-indices, the third one being given by Eq. \eqref{sowEq}. Other possible $q$-indices, corresponding to other properties, could in principle be obtained by using the present $q$-transformations. In contrast, for the present observations for which $q_1 \ne 0$ quite neatly, it is allowed to think that yet unobserved values of $q$ remain to be included within $q$-quadruplets, or even within higher-order cycles. Something similar could be applicable for the present two-term examples for which, once again, $q_1$ neatly differs from zero. Further progress along the present lines will naturally be very welcome.

\end{document}